\documentclass[twocolumn,showpacs,preprintnumbers,amsmath,amssymb]{revtex4}

\usepackage{graphicx}
\usepackage{dcolumn}
\usepackage{bm}

\begin{document}


\title{Exact Effective action for (1+1)-dimensional fermions in an Abelian background at finite temperature and chemical potential}


\author{Soraya G. Maciel}
\author{Silvana Perez}
\affiliation{Faculdade de F\'{\i}sica, Universidade Federal do Par\'a, Bel\'em, Par\'a 66075-110, Brazil}

\date{\today}

\begin{abstract}
In this paper we study the effects of a nonzero chemical potential in the effective action  for massless fermions in $(1+1)$ dimensions in an abelian gauge field background at finite temperature. We calculate  the n-point function and show that the structure of the amplitudes corresponds to a generalization of the structure noted earlier in a calculation without a chemical potential (the associated integrals carry the dependence on the chemical potential). Our calculation shows that the chiral anomaly is unaffected by the presence of a chemical potential at finite temperature. However, unlike the earlier calculation (in the absence of a chemical potential) odd point functions do not vanish. We trace this to the fact that in the presence of a chemical potential the generalized charge conjugation symmetry of the theory allows for such amplitudes. In fact, we find that all the even point functions are even functions of $\mu$ while the odd point functions are odd functions of $\mu$ which is consistent with this generalized charge conjugation  symmetry. We show that the origin of the structure of the amplitudes is best seen from a formulation of the theory in terms of left and right handed spinors. The calculations are also much simpler in this formulation and it clarifies many other aspects of the theory. 
\end{abstract}

\maketitle

\section{\label{intro}Introduction}
 
Thermal effects in quantum field theories have been studied vigorously in the past several years and this has led to a better understanding of such systems \cite{kapusta,bellac,das:book97}. For instance, we now know that new branch cuts develop at finite temperature because of which thermal amplitudes become non-analytic at the origin in the energy-momentum plane \cite{das:book97, non-anal}. This behavior is related with interesting physical phenomena such as screening and plasma oscillations \cite{kapusta, screen-4d, scren-3d}. The effect of temperature on chiral anomalies has also been analysed and it is known  that since anomalies arise as a consequence of ultraviolet divergences at zero temperature and since there is no ultraviolet divergence at finite temperature, there is no finite temperature contribution to the chiral anomaly \cite{das:book97, anomaly}. However, systems at finite temperature and chemical potential (density) have not yet been investigated as vigorously in spite of their many possible applications to physical processes \cite{bellac}.

Calculations of amplitudes at finite temperature are in general more involved than at zero temperature.
However, in an earlier paper \cite{adilson} the complete effective action for massless fermions in an abelian background (which we will refer to as the photon for simplicity) in $(1+1)$ dimensions has been calculated at finite temperature. It was shown there that at finite temperature all the even point functions are nonzero although at zero temperature the effective action is only quadratic in the background fields \cite{karev}. In particular, the nonvanishing (even) n-point amplitudes were found to have a very simple structure of the form

\begin{eqnarray}\label{geral}
\Gamma^{\mu_{1}\mu_{2}\cdots (T)}_{a_{1}a_{2}\cdots}(p_{1},p_{2}, \cdots)&\propto& \left[\delta(p_{1-})\delta(p_{2-})\cdots \times u_-^{\mu_{1}}u_-^{\mu_{2}}\cdots\right] \nonumber \\
&+& \left[\delta(p_{1+})\delta(p_{2+})\cdots \times u_+^{\mu_{1}}u_+^{\mu_{2}}\cdots\right]\nonumber\\
& &
\end{eqnarray}

\noindent with the light-cone variables for any four vector $A^{\mu}$ defined in terms of the null vectors $u^{\mu}_{\pm}$ as

\begin{equation}
u^{\mu}_{\pm}\equiv (1, \mp 1),\qquad A_{\pm} = A^{0}\pm A^{1} = A\cdot u_{\pm},\label{lightconevariables}
\end{equation}

\noindent and we note that the null vectors satisfy the algebraic properties
\begin{equation}
\eta_{\mu\nu} u^{\mu}_{\pm} u^{\nu}_{\pm} = 0,\quad \eta_{\mu\nu} u^{\mu}_{\pm} u^{\nu}_{\mp} = 2.\label{uproperties}
\end{equation}

\noindent The momenta $p_{1},p_{2},\cdots$ in \eqref{geral} denote the independent external momenta for the amplitude while the indices $a_{i}$  represent thermal indices which take the values $\pm$ in the closed time path formalism (which we will also be using, for review see \cite{das:book97}). Besides this simple structure, the amplitudes were shown to be proportional to integrals involving distribution functions whose forms generalize easily for any amplitude.

In this paper, we generalize the results of \cite{adilson} and evaluate the complete effective action for massless fermions in an abelian background in (1+1) dimensions at finite temperature and nonzero chemical potential. We explicitly evaluate the n-point functions and find that while a nonzero chemical potential does not modify the zero temperature amplitudes of the theory, it does lead to new contributions at finite temperature. The structure of the amplitudes at finite temperature and nonzero chemical potential is a generalization of Eq. (\ref{geral}) with the nontrivial dependence on the chemical potential contained in the integrals. In the presence of a chemical potential, the amplitudes with an odd number of photon lines no longer vanish, but continue to have the general structure in \eqref{geral}. We find that the origin of this special structure for the amplitudes is best understood from the Ward identities when the theory is expressed in terms of one dimensional spinors. Our analysis also shows that the chiral anomaly is unchanged in the presence of a chemical potential. On the one hand, this result conforms to the fact that a nonzero chemical potential does not lead to ultraviolet divergence and, therefore, should not modify the chiral anomaly. On the other hand, the response of the particles and antiparticles to a chemical potential is different and would have led one to expect otherwise. This puzzle is again clarified in terms of the one dimensional spinors. 

The paper is organized as follows. In Sec. \ref{sec-2} we present the model and discuss its main properties where we also give the propagators for the theory at finite temperature and chemical potential in the closed time path formalsim \cite{tor:06}. In Sec. \ref{sec-3} we evaluate explicitly the two, three and the four-point functions at finite temperature and chemical potential. We show that they reduce to the results in \cite{adilson} for $\mu \rightarrow 0$. We also show that there is no contribution to the chiral anomaly from the chemical potential. In Sec. \ref{sec-4} we give an alternative method for evaluating these amplitudes in terms of left and right handed spinor fields. The calculations in this formulation are quite simple and clarify the origin of the structures as in \eqref{geral} through the Ward Identities of the theory. This also leads to a better understanding of why there is no new contributions to the chiral anomaly coming from the chemical potential. We end with a brief conclusion in Sec. \ref{sec-5}.

\section{\label{sec-2} The model}

We are considering a theory of massless fermions in the background of an Abelian gauge field in $(1+1)$ dimensions at finite temperature and nonvanishing chemical potential. The Lagrangian density for the theory is given by (we assume $\mu>0$ for simplicity)

\begin{equation}\label{lag}
{\cal L}= i \bar{\psi} \gamma^{\mu} (\partial_{\mu} + i e A_{\mu})\psi +
\mu \bar{\psi}\gamma^0 \psi,
\end{equation}

\noindent where in $(1+1)$ dimensions we use the representation for the Dirac  matrices given by the two dimensional Pauli matrices as 

\begin{equation}
\gamma^0=\sigma_1, \qquad \gamma^1=-i \sigma_2, \qquad \gamma_5 = \sigma_3,\label{gamma}
\end{equation}

\noindent and the metric has the diagonal form $\eta^{\mu \nu}=\mbox{diag}(1,-1)$. From \eqref{gamma} it is easy to see that the gamma matrices satisfy the identity

\begin{equation}
\gamma^{\mu}\gamma^{\nu}= \eta^{\mu\nu}+ \epsilon^{\mu\nu}\gamma_5,\quad \mu,\nu=0,1.\label{gammaidentity}
\end{equation}

\noindent As a result of the identity \eqref{gammaidentity}, the trace of a product of an even number of Dirac matrices is easily seen to take the form

\begin{eqnarray}
\lefteqn{{\rm Tr}\ \big(\gamma^{\mu_{1}}{A\!\!\!\slash}_{1} \gamma^{\mu_{2}}{A\!\!\!\slash}_{2}\cdots \gamma^{\mu_{n}}{A\!\!\!\slash}_{n}\big)}\nonumber\\
& = &\!\!\!A_{1-}\cdots A_{n-} u^{\mu_{1}}_{+}\cdots u^{\mu_{n}}_{+}\! +\! A_{1+}\cdots A_{n+} u^{\mu_{1}}_{-}\cdots u^{\mu_{n}}_{-}\!,\label{identity}
\end{eqnarray}

\noindent where the light-cone components and the null vectors are defined in \eqref{lightconevariables}. As we will see, this identity is very useful in carrying out the calculation of the effective action. We also note here that the theory \eqref{lag} is invariant under a generalized charge conjugation symmetry defined by

\begin{equation}
\psi \rightarrow C\overline{\psi}^{\rm T},\quad A_{\mu}\rightarrow - A_{\mu},\quad \mu\rightarrow -\mu,\label{chargeconjugation}
\end{equation}

\noindent where $C$ denotes the charge conjugation matrix. The fact that the chemical potential changes sign under this generalized charge conjugation symmetry transformation is consistent with the fact that a chemical potential can also be viewed as a constant electrostatic potential.

The effective action  for the photons (Abelian gauge fields) is obtained by integrating out the fermion fields in the path integral and leads to (up to an unimportant  normalization constant)

\begin{equation}\label{geral2}
\Gamma[A]=-i\mbox{Tr} \ln \big(1 -e S(i\partial\!\!\!\slash) A\!\!\!\slash\big),
\end{equation}

\noindent where $S$ denotes the Feynman propagator for the theory in the presence of a chemical potential which in momentum space has the form

\begin{equation}
iS(p) = \frac{i}{p\!\!\!\slash + \mu \gamma^0 + i \epsilon},\label{prop-free}
\end{equation}

\noindent and ``Tr" stands for the trace over the Dirac matrices as well as for the trace over a complete basis of states. At zero temperature, the effective action for the gauge fields can be evaluated in closed form and  contains only quadratic terms in the gauge fields. This result has been used extensively in solving many two dimensional models \cite{sol-2d}. We note here that while this result in the absence of a chemical potential at zero temperature is quite well known, the fact that it holds even in the presence of a nonzero chemical potential is best seen in the mixed space \cite{tor:05, tor:06}, where the dependence of the propagator on the chemical potential is given by an overall phase which gives unity around a closed loop.

However, as explained in \cite{adilson}, at finite temperature the effective action \eqref{geral2} does not have a closed form (even in the absence of a chemical potential). In this case, one has to evaluate the effective action order by order in perturbation theory and we note here the form of the propagator at finite temperature and chemical potential. As mentioned earlier, we will work in the real time formalism known as the closed time path formalism \cite{das:book97} where the propagator has a $2\times 2$ matrix structure $S_{ab}^{\beta,\mu}$ with $a,b =\pm$ denoting the thermal indices corresponding to the doubled degrees of freedom in the real time formalisms. For completeness we note here the explicit forms of the components of the propagator

\begin{eqnarray}
S_{++}^{(\beta, \mu) }(p) &=& (p\!\!\!\slash+ \mu \gamma^0)[\frac{1}{(p_0+\mu)^2- \vec{p}^2 + i
    \epsilon}\nonumber \\
&+& 2 \pi i n_F(p_0{\rm sgn}(p_0+\mu))\delta((p_0+\mu)^2-
    {\vec{p}}^2) ],\nonumber\\
S_{+-}^{(\beta, \mu) }(p)&=&2 \pi i (p\!\!\!\slash+ \mu \gamma^0)\delta((p_0+\mu)^2-
    {\vec{p}}^2)\nonumber\\
&\times& [- \theta(-p_0- \mu)+
    n_F(p_0{\rm sgn}(p_0+\mu))],\label{prop}\\
S_{-+}^{(\beta, \mu) }(p)&=&2 \pi i (p\!\!\!\slash+ \mu \gamma^0)\delta((p_0+\mu)^2-
    {\vec{p}^2})\nonumber\\
&\times& [- \theta(p_0+ \mu)+
    n_F(p_0{\rm sgn}(p_0+\mu))], \nonumber\\
S_{--}^{(\beta, \mu) }(p) &=& (p\!\!\!\slash+ \mu \gamma^0)[\frac{-1}{(p_0+\mu)^2- \vec{p}^2 - i
    \epsilon}\nonumber \\
&+&\!\! 2 \pi i n_F(p_0{\rm sgn}(p_0+\mu))\delta((p_0+\mu)^2-
    {\vec{p}^2}) ],\nonumber
\end{eqnarray}

\noindent where $n_{F}$ denotes the Fermi-Dirac distribution function. As is clear from the forms of the propagator in \eqref{prop}, each component of the propagator  is the sum of a zero temperature part and a part that explicitly depends on temperature which is rather useful in calculating the temperature dependent part of an amplitude. 

\section{\label{sec-3}N-point functions}

As we have mentioned, we do not expect the effective action at finite temperature and chemical potential to have a closed form. Therefore, let us start by calculating a few low order amplitudes to see if there is a general pattern for all the higher point amplitudes. We will carry out the calculation only for the amplitudes with $+$ vertices since the other ones will be quite similar.

\subsection{Two-point function}

The simplest nontrivial calculation is that for the photon two point function $\Gamma^{\mu\nu(\beta,\mu)}_{++}$ arising from the fermion loop diagram which in the momentum space has the form
\begin{widetext}
\begin{eqnarray}\label{eq.2}
i \Gamma^{\mu\nu(\beta, \mu)}_{++}(p) &=& - e^2 \int \frac{d^2 k}{(2
  \pi)^2} \mbox{Tr}\gamma^{\mu}S_{++}^{(\beta, \mu)
  }(k+p)\gamma^{\nu}S_{++}^{(\beta, \mu) }(k) = - e^2 \int \frac{d^2 k}{(2
  \pi)^2} \mbox{Tr}\big(\gamma^{\mu}(\bar{k}\!\!\!\slash+p\!\!\!\slash)\gamma^{\nu}\bar{k}\!\!\!\slash\big) \nonumber \\
&& \quad \times \left [\frac{1}{(\bar{k})^2} + 2 \pi i
  n_F(k_0\,{\rm sgn}(\bar{k}_0))\delta(\bar{k}^2) \right]\!\! \left[\frac{1}{(\bar{k}+p)^2} + 2 \pi i
  n_F((k_0+p_0)\,{\rm sgn}(\bar{k}_0+p_0)\delta((\bar{k}+p)^2)\right],
\end{eqnarray}
\end{widetext}

\noindent where, for simplicity, we have identified a new four-vector $\bar{k}_{\mu}$ with

\begin{equation}
\bar{k}_{\mu}= (k_0+\mu,-\vec{k}).\label{bark}
\end{equation}

\noindent We note here that even though we do not show it explicitly for simplicity, the Feynman $i\epsilon$ prescription in the zero temperature propagators is understood.The zero temperature part of the two point function is easily seen from \eqref{eq.2} to come from the product of the two zero temperature propagators and needs to be regularized. However, the explicitly temperature dependent part of the two point function which we denote by $i \Gamma_{++}^{\mu \nu(T, \mu)}(p)$ (and which is obtained from \eqref{eq.2} by subtracting out the zero temperature part) is well behaved. Using \eqref{identity} the trace over the Dirac matrices in \eqref{eq.2} can be easily done. Furthermore, the integration over $k_{0}$ using the delta function is also straightforward and with some algebra, the temperature dependent photon two point function is obtained to be 

\begin{eqnarray}\label{final2}
i \Gamma_{++}^{\mu \nu(T, \mu)}(p)&=&
-\frac{e^2}{4}[\delta(p_-)u_-^{\mu}u_-^{\nu} +
  \delta(p_+)u_+^{\mu}u_+^{\nu} ] \nonumber \\ 
& &\times \int dk^1
\mbox{sgn}(k^1) \mbox{sgn}(k^1+ p^1)I_2^{(T,\mu)} \nonumber \\
\end{eqnarray}

\noindent  where

\begin{eqnarray}\label{i2}
I_2^{(T,\mu)} &=&  \big[ \big( n_F((k^1 -\mu)\mbox{sgn}(k^1)) \nonumber \\
& & +  n_F((k^1 + p^1 -\mu)\mbox{sgn}(k^1+ p^1)) \nonumber \\
& &  - 2 n_F((k^1 - \mu)\mbox{sgn}(k^1)) \nonumber \\
& & \times n_F((k^1 + p^1 -
  \mu)\mbox{sgn}(k^1+ p^1))\big) \nonumber \\
& & + (\mu\rightarrow -\mu)\big].\nonumber\\
& &
\end{eqnarray}

There are several things to note here. First, the structure of the two point function coincides with \eqref{geral} and that in the limit $\mu\rightarrow 0$, the integrand \eqref{i2} reduces to that obtained in  \cite{adilson}. Furthermore, even in the presence of a nonzero potential, we note from Eq.(\ref{final2}) that
the two point function is manifestly transverse, namely,

\begin{eqnarray}
p_{\mu}  \Gamma_{++}^{\mu \nu(T, \mu)} (p)&\propto& p_{\mu}[\delta(p_-)u_-^{\mu}u_-^{\nu} +
  \delta(p_+)u_+^{\mu}u_+^{\nu} ] \nonumber \\
&=& \delta(p_-)p_- u_-^{\nu} + \delta(p_+) p_+ u_+^{\nu}\nonumber \\
&\equiv& 0,
\end{eqnarray}

\noindent where we have used \eqref{lightconevariables} to write 

\begin{equation}
p \cdot u_{\pm} = p_{\pm}.
\end{equation}

We note here that in $(1+1)$ dimensions the chiral two point function $\Gamma^{\mu\nu}_{5,++}(p)$ (where one of the vertices is, say, $\gamma_{5}\gamma^{\mu}$) can be obtained from the photon two point function because of gamma matrix identities. Thus, the temperature dependent chiral two point function can be obtained from \eqref{final2} to correspond to 

\begin{equation}
\Gamma_{5,++}^{\mu \nu (T, \mu)} (p) =\epsilon^{\mu}_{\lambda}
\Gamma_{++}^{\lambda \nu (T,\mu)} (p).
\end{equation}

\noindent The structure of the chiral two point function now follows to be (the integral and the multiplicative factors remain the same)

\begin{eqnarray}
\Gamma_{5,++}^{\mu \nu (T, \mu)} (p) &\propto& \epsilon^{\mu \lambda} [\delta(p_-) u_{- \lambda
  }u_-^{\nu} + \delta(p_+) u_{+ \lambda}u_+^{\nu}] \nonumber \\
&\equiv& - \delta(p_-) u_-^{\mu}u_-^{\nu} + \delta(p_+) u_+^{\mu}u_+^{\nu},
\end{eqnarray}

\noindent where we have used the identity 

\begin{equation}
\epsilon^{\mu \lambda} u_{\pm \lambda} = \pm u^{\mu}_{\pm}.
\end{equation}

\noindent As a result, we obtain

\begin{eqnarray}
p_{\mu}\Gamma_{5,++}^{\mu \nu (T, \mu)} (p) &\propto& - \delta(p_-)p_- u_-^{\nu} + \delta(p_+)p_+
  u_+^{\nu}\nonumber \\ 
&\equiv& 0.
\end{eqnarray}

\noindent This shows that there is no contribution to the chiral anomaly at finite temperature and nonzero chemical potential. The absence of a temperature dependent contribution to the chiral anomaly is well known and the discussion in the next section will clarify why the chemical potential does not modify the anomaly.

Using the other components of the fermion propagator in \eqref{prop} we can also evaluate the other
components of the photon self-energy at finite temperature and chemical
potential. For example, for the temperature dependent part of the $+-$ component of the two point function we have (see \cite{das:book97} for a definition of this as well as retarded and advanced functions)  

\begin{widetext}
\begin{eqnarray}
i \Gamma_{+-}^{\mu \nu (T, \mu)} (p) &=& \frac{e^{2}}{4}\big[2(- \delta(p_{-})u^{\mu}_{-}u^{\nu}_{-} + \delta(p_{+}) u^{\mu}_{+}u^{\nu}_{+}) \int dk^{1}\,{\rm sgn} (k^{1})\big(n_{F} ((k^{1}+\mu){\rm sgn}(k^{1})) + n_{F} ((k^{1}-\mu){\rm sgn}(k^{1}))\big)\nonumber\\
& &\qquad + (\delta(p_{-})u^{\mu}_{-}u^{\nu}_{-} + \delta (p_{+}) u^{\mu}_{+}u^{\nu}_{+}) \int dk^{1}\,{\rm sgn}(k^{1}){\rm sgn}(k^{1}+p^{1})\,I_{2}^{(T,\mu)}\big] = - i\Gamma_{++}^{\mu \nu (T, \mu)} (p),
\end{eqnarray}
\end{widetext}

\noindent where we have used the definition in \eqref{i2} as well as the fact that the first integral vanishes because of anti-symmetry. It follows now that the temperature dependent part of the retarded two point function in the presence of a nonzero chemical potential identically vanishes, namely,

\begin{equation}
i \Gamma_R^{\mu \nu (T,\mu)} \equiv i \Gamma_{++}^{\mu \nu(T,\mu)} + i \Gamma_{+-}^{\mu \nu(T,\mu)} = 0.
\end{equation}

\noindent The advanced two point function also can be seen to vanish in a similar manner. Such a behavior was already noted in the absence of a chemical potential in \cite{adilson} and this raises an interesting open question as to how the nontrivial structures for the Feynman amplitudes found in this model can be obtained in the imaginary time formalism. 

\subsection{\label{3point} Higher point functions}

Let us next calculate the three point function for the photon at finite temperature and chemical potential. An amplitude with an odd number of external photon lines generally vanishes at zero temperature as well as at finite temperature (in the absence of a chemical potential) by charge conjugation (Furry's theorem). However, as we will see this is not true when there is a nonzero chemical potential. The three point amplitude will involve two graphs where two of the external photon lines are exchanged. Denoting the independent momenta of the external lines to be $(p,q)$ associated with photon lines with indices $(\nu,\lambda)$ respectively, the three point function has the form

\begin{widetext}
\begin{eqnarray}\label{eq.33}
i \Gamma_{+++}^{\mu \nu \lambda(\beta, \mu)}(p,q) &=& - e^3 \int
\frac{d^2k}{(2 \pi)^2}\Big[{\rm Tr}\big(\gamma^{\mu} \bar{k}\!\!\!\slash
\gamma^{\nu}(\bar{k}\!\!\!\slash+p\!\!\!\slash)\gamma^{\lambda}(\bar{k}\!\!\!\slash+p\!\!\!\slash+
q\!\!\!\slash)\big)\left [\frac{1}{(\bar{k})^2} + 2 \pi i
  n_F(k_0{\rm sgn}(\bar{k}_0))\delta(\bar{k}^2) \right]\nonumber\\
&\times&\left [\frac{1}{(\bar{k}+p)^2} + 2 \pi i
  n_F((k_0+p_{0}){\rm sgn}(\bar{k}_0+p_{0}))\delta((\bar{k}+p)^2) \right]\nonumber\\
 &\times& \!\!\!\!\left [\frac{1}{(\bar{k}+p+q)^2} + 2 \pi i
  n_F((k_0+p_{0}+q_{0}){\rm sgn}(\bar{k}_0+p_{0}+q_{0}))\delta((\bar{k}+p+q)^2) \right] \!\!+ (p \leftrightarrow q), (\nu\leftrightarrow\lambda)\Big]\!.
\end{eqnarray}
\end{widetext}

\noindent Using \eqref{identity} the Dirac trace can be simplified and carrying out the integration over $k_{0}$ using the delta function, the temperature dependent part of the three point function in the presence of a nonzero chemical potential is obtained to be

\begin{widetext}
\begin{eqnarray}
i \Gamma^{\mu \nu \lambda(T,\mu)}_{+++}(p,q) &=& \frac{i\pi e^3}{4}
[\delta(p_-)\delta(q_-)u_-^{\mu}u_-^{\nu}u_-^{\lambda} +
 \delta(p_+)\delta(q_+)u_+^{\mu}u_+^{\nu}u_+^{\lambda} ]\int dk^1 \mbox{sgn}(k^1)\mbox{sgn}(k^1+ p^1)\mbox{sgn}(k^1+
 P^1_{T(3)}) I_3^{(T,\mu)} \nonumber \\
& & + \mbox{all permutations of external momenta}
\end{eqnarray}
\end{widetext}

\noindent where we have identified $P^{1}_{T(3)}=p^{1}+q^{1}$ for simplicity and the integrand  $I_3^{(T,\mu)}$ is defined to be  

\begin{widetext}
\begin{eqnarray}
I_3^{(T,\mu)}&=&\big[(n_F((k^1  - \mu)\mbox{sgn}(k^1)) + n_F((k^1 +p^1 - \mu)\mbox{sgn}(k^1+p^1)) + n_F((k^1 +P^1_{T(3)} - \mu) \mbox{sgn}(k^1+P^1_{T(3)})) \nonumber \\
& & - 2n_F((k^1  - \mu)\mbox{sgn}(k^1)) n_F((k^1 +p^1 - \mu)\mbox{sgn}(k^1+p^1)) \nonumber \\
& & - 2 n_F((k^1  - \mu)\mbox{sgn}(k^1))n_F((k^1+P^1_{T(3)}  -
 \mu)\mbox{sgn}(k^1+P^1_{T(3)})) \\
& & - 2 n_F((k^1 +p^1 - \mu)\mbox{sgn}(k^1+p^1)) n_F((k^1+P^1_{T(3)}  - \mu)\mbox{sgn}(k^1+P^1_{T(3)}))\nonumber \\
& & + 4n_F((k^1  - \mu)\mbox{sgn}(k^1)) n_F((k^1+p^1  -
 \mu)\mbox{sgn}(k^1+p^1)) n_F((k^1+P^1_{T(3)}  - \mu)\mbox{sgn}(k^1+P^1_{T(3)}))) - (\mu\rightarrow -\mu)\big] \nonumber.\label{i3}
\end{eqnarray}
\end{widetext}

\noindent There are several things to note from this calculation. First, in the limit of vanishing chemical potential (note the last term $- (\mu \rightarrow - \mu)$ in \eqref{i3}), the three point function vanishes  \cite{adilson} as we would expect from charge conjugation invariance. In the presence of a chemical potential, however, the generalized charge conjugation invariance of the model \eqref{chargeconjugation} does allow for odd functions of $\mu$ for the odd point functions of the theory and this is exactly what we see here.

Let us next calculate the four point function of the theory. Without going into technical details (which are similar to what we have already discussed), we note here that when the graphs with all the possible permutations are added, the temperature dependent part of the four point function takes the form (we choose $p,q,l$ to denote the independent momenta of the four point function)
\begin{widetext}
\begin{eqnarray}
i \Gamma^{\mu \nu \lambda\sigma (T,\mu)}_{++++}(p,q,l) &=& \frac{\pi^2 e^4}{4}\big(\delta(p_-)\delta(q_-)\delta(l_-)u_-^{\mu}u_-^{\nu}u_-^{\lambda}u_-^{\sigma} +
 \delta(p_+)\delta(q_+)\delta(l_+)u_+^{\mu}u_+^{\nu}u_+^{\lambda}u_+^{\sigma}\big)\nonumber \\
 & & \times \int dk^1 \mbox{sgn}(k^1)\mbox{sgn}(k^1+ p^1) \mbox{sgn}(k^1+
 P^1_{T(3)}) \mbox{sgn}(k^1+ P^1_{T(4)})  I_4^{(T,\mu)}\nonumber \\
& & + \mbox{all permutations of external momenta}
\end{eqnarray}
\end{widetext}

\noindent where $P^{1}_{T(4)} = p^1+q^1+l^1$ and $I_4^{(T,\mu)}$ is defined as  
\begin{widetext}
\begin{eqnarray}
I_4^{(T,\mu)} &=& \Big[\Big((n_F((k^1-\mu)\mbox{sgn}(k^1)) + n_F((k^1+p^1-\mu)\mbox{sgn}(k^1+ p^1)) + \mbox{ linear permutations})   \nonumber \\
& & - 2 (n_F((k^1-\mu)\mbox{sgn}(k^1))n_F((k^1+p^1-\mu)\mbox{sgn}(k^1+ p^1)) + \mbox{quadratic permutations}) \nonumber \\
& & + 4 (n_F((k^1-\mu)\mbox{sgn}(k^1))n_F((k^1+ p^1-\mu)\mbox{sgn}(k^1+p^1))n_F((k^1+ P^1_{T(3)}-\mu)\mbox{sgn}(k^1+P^1_{T(3)}))  \nonumber\\
& & \qquad + \mbox{cubic permutations}) \nonumber \\
& & - 8 n_F((k^1-\mu)\mbox{sgn}(k^1))n_F((k^1+p^1-\mu)\mbox{sgn}(k^1+p^1))n_F((k^1+P^1_{T(3)}-\mu)\mbox{sgn}(k^1+P^1_{T(3)}))\nonumber \\
&& \qquad \times n_F((k^1+P^1_{T(4)}-\mu)\mbox{sgn}(k^1+P^1_{T(4)}))\Big)  + \Big(\mu\rightarrow -\mu\Big)\Big].
\end{eqnarray}
\end{widetext}

We can continue to calculate the general structure of the $n$-point
functions at finite temperature and chemical potential. However, a pattern for higher point functions is already emerging. We  see that the structure of the amplitudes always decomposes  into two parts, one with only products  of the type $u_+ \delta(p_+) $ and other with products of  the form $u_-\delta(p_-)$ (which is a generalization of the $\mu=0$ result \cite{adilson}). The associated integrals also have a simple generalization as we go to higher point functions so that for the temperature dependent part of the $n$-point function we have

\begin{widetext}
\begin{eqnarray}\label{eq.66}
i \Gamma^{\mu_1 \cdots \mu_n (T,\mu)}_{+\cdots +}(p_1,\cdots ,p_{n-1}) &=& \frac{(-i e)^{n}(\pi)^{n-2}}{4}[\delta(p_{1-})\cdots \delta(p_{(n-1)-})u_-^{\mu_1}\cdots u_-^{\mu_n} + \delta(p_{1+})\cdots \delta(p_{(n-1)+})u_+^{\mu_1}\cdots u_+^{\mu_n} ]\nonumber \\
 & & \times \int dk^1 \mbox{sgn}(k^1)\mbox{sgn}(k^1+ p^1) \mbox{sgn}(k^1+ P^1_{T(3)})\cdots \mbox{sgn}(k^1+ P^1_{T(n)})I_n^{(T,\mu)} \nonumber \\
& & + \mbox{all permutations off external momenta.}
\end{eqnarray}
\end{widetext}

\noindent where $P^{1}_{T(n)} = p_{1}^1+p_{2}^1+\cdots +p_{n-1}^1$ and  $I_n^{(T,\mu)}$ is given by

\begin{widetext}
\begin{eqnarray}
\lefteqn{I_n^{(T,\mu)} = \Big[\Big( [n_F((k^1-\mu)\mbox{sgn}(k^1)) + n_F((k^1+p^1_{1}-\mu)\mbox{sgn}(k^1+ p^1_{1})) + \mbox{linear permutations}] }  \nonumber \\
&-&2 [n_F((k^1-\mu)\mbox{sgn}(k^1))n_F((k^1+p^1_{1}-\mu)\mbox{sgn}(k^1+ p^1_{1})) + \mbox{quadratic permutations}] \nonumber \\
&+& 2^{2} [n_F((k^1-\mu)\mbox{sgn}(k^1))n_F((k^1+ p^1_{1}-\mu)\mbox{sgn}(k^1+p^1_{1}))n_F((k^1+p^1_{1}+p^1_{2}-\mu)\mbox{sgn}(k^1+p^{1}_{1}+p^{1}_{2})) + \mbox{cubic permutations}] \nonumber \\
&+& \cdots \Big)  + (-1)^{n} \Big(\mu\rightarrow -\mu\Big)\Big].
\end{eqnarray}
\end{widetext}

\noindent The symmetry (anti-symmetry) of all the amplitudes under $\mu\rightarrow -\mu$ is manifest. In the limit of vanishing chemical potential all the odd point amplitudes vanish and the even point amplitudes reduce to those already calculated in \cite{adilson}. The effective action for the photons now follows from the explicit form of the amplitudes in a straightforward manner. For example, the effective action at the $n$th order is simply given by

\begin{equation}
\Gamma^{(n)}_{+\cdots +}[A] = \frac{1}{n!} \int \prod_{i=1}^{n}\frac{d^{2}p_{i}}{(2\pi)^{2}} A_{\mu_{1}}\cdots A_{\mu_{n}} \Gamma^{\mu_{1}\cdots \mu_{n}}_{+\cdots +}.
\end{equation}

\section{\label{sec-4} Alternative derivation of the results}

The particular structure of the amplitudes in \eqref{eq.66} involving products of light-cone delta functions as well as light-cone (null) vectors is quite puzzling. In some sense it is reminiscent of amplitudes in some $0+1$ dimensional models where the amplitudes involve delta functions of energy \cite{dunne,barcelos}. In order to better understand the structure of the $n$-point functions in this model, therefore, we present an alternative derivation in terms of one dimensional fermions which also clarifies various other features of our results. Let us note that if we decompose the fermions in the theory into  left and right handed components as 

\begin{eqnarray}
\psi_L & = & P_L \psi = \frac{1}{2} (1-\gamma_5) \psi,\nonumber\\
\psi_R & = & P_R \psi = \frac{1}{2}(1+\gamma_5)\psi,\label{LRfermions}
\end{eqnarray} 

\noindent then in $(1+1)$ dimensions these fermions effectively become one dimensional. The Lagrangian density \eqref{lag} can be written in terms of these components as 

\begin{equation}\label{lagrangian}
{\cal L} =  i \psi^{\dagger}_R (\partial_- - i \mu + i e A_-)\psi_R  + i \psi^{\dagger}_L (\partial_+ - i \mu + i e A_+)\psi_L,
\end{equation}

\noindent where we have used the notation in \eqref{lightconevariables} to write

\begin{equation}
\partial_{\pm}\equiv \partial\cdot u_{\pm} = \partial_0 \mp \partial_1, \qquad A_{\pm}\equiv A\cdot u_{\pm} = A_0 \mp A_1.
\end{equation}

There are several things to note from the structure of \eqref{lagrangian}. The Lagrangian density of the theory effectively becomes a sum of two decoupled Lagrangian densities for one dimensional left and right moving fermions interacting with independent light-cone components of the gauge field. Therefore, we suspect that the amplitudes will involve only the $+$ components or the $-$ components of the photon field. Of course, such an argument will not hold if there are divergence that need to be regularized which may introduce mixing between the $+$ and $-$ photon components. This is indeed what happens at zero temperature where we have the chiral anomaly which generates a mixed photon two point function with $+$ and $-$ components. However, at finite temperature there is no ultraviolet divergence that needs to be regularized and, therefore, if we are looking only at the temperature dependent parts of the amplitudes, there will be no mixing between the $+$ and $-$ photon components. This is an alternative way to see why the chiral anomaly will not receive any temperature dependent correction. In fact, let us note here that while the chemical potential distinguishes
between the positive and the negative energy states, it is clear from 
Eq.(\ref{lagrangian}) that it treats the left and right handed components of the fermion fields 
symmetrically. As a result, the chemical potential does not
contribute to the chiral anomaly which arises from a difference between the left and the right handed fermions (zero modes).

At finite temperature and a nonzero  chemical potential, the propagators for these component fermions can be easily derived to have the form (to avoid confusion between the light-cone components and the thermal indices, in this section we will avoid using the thermal indices keeping in mind that we are going to be looking at amplitudes with only $+$ thermal index for simplicity)

\begin{widetext}
\begin{eqnarray}
S_{R}^{(\beta,\mu)} (p)&=& \bar{p}_+ P_R \Big[\frac{1}{\bar{p}_+\bar{p}_- + i\epsilon} + 2 \pi i n_F(p_0 \mbox{sgn}(\bar{p}_0)) \delta(\bar{p}_+\bar{p}_-)\Big] = P_R \Big[ \frac{1}{\bar{p}_- + i \epsilon \mbox{sgn}(\bar{p}_+)} + 2 \pi i n_F(p_0 \mbox{sgn}(\bar{p}_0)) \mbox{sgn}(p^1)\delta(\bar{p}_-)\Big],\nonumber\\
S_{L}^{(\beta,\mu)}(p)&=& \bar{p}_- P_L \Big[\frac{1}{\bar{p}_+\bar{p}_- + i\epsilon} + 2 \pi i n_F(p_0 \mbox{sgn}(\bar{p}_0)) \delta(\bar{p}_+\bar{p}_-)\Big] = P_L \Big[ \frac{1}{\bar{p}_+ + i \epsilon \mbox{sgn}(\bar{p}_-)} - 2 \pi i n_F(p_0 \mbox{sgn}(\bar{p}_0)) \mbox{sgn}(p^1)\delta(\bar{p}_+)\Big],\nonumber\\
& &\label{componentpropagators}
\end{eqnarray}
\end{widetext}

\noindent where the projection operators $P_{L},P_{R}$ are defined in \eqref{LRfermions} and as in \eqref{bark} we have identified

\begin{equation}
\bar{p}_{\pm} = p^{0}+\mu \pm p^{1}.
\end{equation}

As we have already argued, at finite temperature the Lagrangian density (\ref{lagrangian}) will give rise to amplitudes only of the types

\begin{eqnarray}
\Gamma_{+++ \cdots +} &=& u_{\mu_{1}+}\cdots u_{\mu_{n}+} \Gamma^{\mu_{1}\cdots \mu_{n}},\nonumber\\
\Gamma_{---\cdots -} &=& u_{\mu_{1}-}\cdots u_{\mu_{n}-} \Gamma^{\mu_{1}\cdots \mu_{n}},\label{Gamma+-}
\end{eqnarray}

\noindent where we have suppressed the thermal indices of the amplitudes to avoid confusion (and we are looking at the amplitudes with only $+$ thermal indices for simplicity).  We note here that using \eqref{uproperties} we can see that the tensor structure of the amplitudes in \eqref{eq.66} already contains this fact that there can be no thermal amplitude with mixed indices, but the origin for this becomes clear from the formulation of the theory as in \eqref{lagrangian} in terms of one dimensional fermions. Calculations with the Lagrangian density \eqref{lagrangian} are extremely simple primarily because there are no tensor structures or nontrivial Dirac matrices to deal with and the propagators are first order. Let us illustrate this with the calculation of the two point function.

We note that the two point function for the $+$ photon components is given by (the thermal indices are $+$ which we are suppressing)

\begin{equation}
i \Gamma_{++}^{(\beta,\mu)} (p) = - e^2 \int \frac{d^2k}{(2 \pi)^2}{\rm Tr} S^{(\beta, \mu)}_{L}(k+p)  S^{(\beta, \mu)}_{L}(k),
\end{equation}

\noindent where the propagator for the left handed fermions is given in \eqref{componentpropagators}. Separating out the zero temperature part, the explicit finite temperature contribution is given by 

\begin{widetext}
\begin{eqnarray}
i \Gamma_{++}^{(T,\mu)} (p) &=& ie^2 \int \frac{d^2k}{(2 \pi)}\Big[\frac{n_F(k_0\mbox{sgn}(\bar{k}_0))\delta(\bar{k}_+)\mbox{sgn}(k^1)}{p_+ + i \epsilon \mbox{sgn}(\bar{k}_- + p_-)}+ \frac{n_F((k_0+p_0)\mbox{sgn}(\bar{k}_0+p_0))\delta((\bar{k} + p)_+) \mbox{sgn}(k^1+p^1)}{-p_+ + i \epsilon \mbox{sgn}(\bar{k}_- )}\nonumber \\
& &\qquad -\ 2 \pi i n_F((k_0)\mbox{sgn}(\bar{k}_0))n_F((k_0+p_0)\mbox{sgn}(\bar{k}_0+p_0)) \delta(\bar{k}_+)\delta(\bar{k}_{+}+ p_+)\mbox{sgn}(k^1)\mbox{sgn}(k^1+p^1)\Big].\label{twopointT}
\end{eqnarray}
\end{widetext}

\noindent where we have used the fact that $P_{L}$ is a projection operator defined in \eqref{LRfermions} and that the trace of the projection operator in $1+1$ dimensions yields unity. Using basic identities such as 

\begin{equation}
\frac{1}{p_+ + i \epsilon {\rm sgn} (\bar{k}_{-}+p_{-})}= \frac{1}{p_{+}} - \pi i {\rm sgn} (\bar{k}_{-}+p_{-})\,\delta (p_+),
\end{equation}

\noindent and the simple symmetry properties of the integral, we can rewrite \eqref{twopointT} as

\begin{equation}
i \Gamma_{++}^{(T,\mu)} (p) = - \frac{e^2}{4} \delta(p_+)\int dk^1 \mbox{sgn}(k^1)\mbox{sgn}(k^1+ p^1) I_2^{\mu} ,
\end{equation}

\noindent where $I_2^{\mu}$ is defined in Eq. (\ref{i2}). On the other hand, recalling the definition in \eqref{Gamma+-} and using the properties in \eqref{uproperties},  we note that this indeed coincides exactly with the $++$ amplitude in  Eq. (\ref{final2}). This simply demonstrates that this alternative method of derivation of the amplitudes is much simpler and can be extended to any $n$-point amplitude in the theory.

\subsection{Ward Identities}

Just as the tensor structure of the theory is easily understood in terms of the one dimensional fermions in \eqref{lagrangian}, the delta function structure is similarly best understood in this description. We note that the Lagrangian density \eqref{lagrangian} has two indenpendent gauge invariances given by

\begin{eqnarray}
\psi_R &\rightarrow& e^{-i \alpha_+(x^+)}\psi_R, \quad A_- \rightarrow A_- + \frac{1}{e} \partial_- \alpha_+(x^+),\label{inv1}\\
\psi_L &\rightarrow& e^{-i \alpha_-(x^-)}\psi_L,\quad A_+ \rightarrow A_+ + \frac{1}{e} \partial_+ \alpha_-(x^-).\label{inv2}
\end{eqnarray}

\noindent These are the independent left and right gauge invariances of the theory at the tree level described by $U(1)\times U(1)$. We know that the chiral anomaly of the theory breaks this symmetry down to a vector $U(1)$ group. Therefore, in general, the Ward identities following from these tree level symmetry would be anomalous. However, as is known and as we have already argued the anomaly is independent of the temperature. Consequently, the temperature dependent part of the amplitudes will satisfy the unbroken Ward identities of $U(1)\times U(1)$.

The Ward identity following from the symmetry transformations in Eq. (\ref{inv1}) leads to (for every coordinate)  

\begin{equation}
\partial_- \Gamma_{---\cdots -}=0,
\end{equation}

\noindent which in momentum space would become

\begin{equation}
p_{i -} \Gamma_{--- \cdots -} (p_{1},p_{2},\cdots)=0,\quad i=1,2,\cdots,\label{deltaconstraint}
\end{equation}

\noindent where the $p_{i}$ represent the independent momenta of the amplitude. Such a relation is quite similar to the Ward identities in the $0+1$ dimensional theories in \cite{dunne,barcelos} and the solution of \eqref{deltaconstraint} leads to

\begin{equation}
\Gamma_{---\cdots} (p_{i}) \propto \delta (p_{1-})\delta (p_{2-})\cdots.
\end{equation}

\noindent In a parallel manner, the Ward identity following Eq. (\ref{inv2}) leads to 

\begin{equation}
\Gamma_{+++\cdots +} (p_{i}) \propto \delta (p_{1+})\delta(p_{2+})\cdots.
\end{equation}

\noindent The Ward identities of the theory, therefore, explain the delta function structure  of the amplitudes. As a result, the formulation in terms of the one dimensional fermions clarifies the origin of the particular structure of the amplitudes in this theory and furthermore, the calculation of the associated integrals are simpler in this framework.

\section{\label{sec-5} Conclusions}

In this paper we have studied systematically the  finite temperature and chemical potential effects in the effective action for (1+1)-dimensional massless fermions in an Abelian gauge field background. We have explicitly evaluated the n-point functions and  found a generalization of a previous calculation \cite{adilson} carried out at finite temperature (without a chemical potential). There is no new contributions to the chiral anomaly in the presence of a chemical potential. However, unlike the earlier calculation where odd point functions vanished because of charge conjugation, in the presence of a chemical potential we find that the odd point functions are all nonvanishing. This is traced to the fact that the generalized charge conjugation symmetry in the presence of such a chemical potential does not prohibit such terms. In fact, this symmetry requires that all the even point functions should be even under $\mu\rightarrow -\mu$ while the odd point functions should be odd (this is another way of seeing how the odd point functions vanish in the absence of a chemical potential). We have calculated explicitly the $n$-point functions where this symmetry is manifest. All the amplitudes (even and odd point functions) continue to have the general structure already found in \cite{adilson}. The origin of this structure becomes clear if we formulate the theory in terms of one dimensional fermions (left and right handed spinors). We have shown that both the tensor structure as well as the delta function structure of the amplitudes follows trivial from the symmetries and the Ward identities of the theory in the formulation in terms of one dimensional fermions which also makes contact with some earlier results in $0+1$ dimensional theories \cite{dunne,barcelos}. We have shown with a simple calculation how the evaluation of the amplitudes also becomes a lot simpler in this formulation and it also sheds light on several other aspects of the theory.

\begin{acknowledgments}
We would like to thank Prof. A. Das for many helpfull discussions. This work was supported by CNPq and CAPES (Brazil).
\end{acknowledgments}

\end{document}